\documentclass[prl,twocolumn,showpacs,amsmath,amssymb,showkeys,nofootinbib]{revtex4}

\usepackage{graphicx}% Include figure files
\usepackage{dcolumn}% Align table columns on decimal point
\usepackage{bm}% bold math
\usepackage{times}

\newcommand{\prm}{\prime}
\newcommand{\mrm}{\mathrm}

\begin{document}

%\preprint{APS/123-QED}

\title{A dynamic nonlinear model for saturation in industrial growth}

\author{Arnab K. Ray}
\email{akr@hbcse.tifr.res.in}
\affiliation{Homi Bhabha Centre for Science Education, TIFR, 
V. N. Purav Marg, Mankhurd, Mumbai 400088, India}

%\date{\today}

\begin{abstract}
A general nonlinear logistic equation has been proposed to model long-time 
saturation in industrial growth. An integral solution of this equation
has been derived for any arbitrary degree of nonlinearity. A time scale 
for the onset of nonlinear saturation in industrial growth can be estimated 
from an equipartition condition between nonlinearity and purely exponential 
growth. Precise 
predictions can be made about the limiting values of the annual revenue 
and the human resource content that an industrial organisation may attain. 
These variables have also been modelled to set up an autonomous first-order
dynamical system, whose equilibrium condition forms a stable node (an
attractor state) in a
related phase portrait. The theoretical model has received close support 
from all relevant data pertaining to the well-known 
global company, {\it IBM}. 
\end{abstract}

\pacs{89.65.Gh, 05.45.-a}
\keywords{Economics, econophysics, business and management; Nonlinear 
dynamics and dynamical systems}
\maketitle

In view of the current economic recession that is prevailing globally,  
it has become imperative to furnish mathematical models of greater 
degrees of quantitative accuracy to understand stagnation in economic 
growth, be it of states or of industrial organisations. When it comes 
to making forecasts about how the future might unfold for a state 
economy or an organisation, one realises the acute necessity for 
proper deterministic models. A recognition 
is gradually gaining currency that economic models should have robust 
predictive power, and this power should issue from the compatibility
of the pertinent mathematical models with empirical data~\cite{bouch}. 

Addressing this requirement is the principal objective of this work,
and this has been done by making a case study on industrial growth.
Many aspects of industrial growth lend themselves to well-formulated
mathematical analyses. The health of a company is to be
judged from the revenue that it is capable of generating, as well
as the extent of human resource that it is capable of employing
in achieving its objectives. One could make quantitative measures
of all these variables, and this makes it relatively easy to have
a clear understanding of industrial growth pattern, as well as to 
posit a mathematical model for it.
The approach to these issues here is predominantly based on the use 
of standard mathematical tools of nonlinear dynamics and dynamical
systems~\cite{braun,stro,js99}. 
There is a general appreciation that even when an industrial
organisation displays noticeable (very commonly exponential) 
growth in the early stages,
there is a saturation of this growth towards a terminal end after
the elapse of a certain scale of time~\cite{aghow}. 
As the system size (reflective of the scale of
operations) begins to grow through the passage of time, a
self-regulatory mechanism takes effective control over growth
and gradually drives the system towards a saturated terminal state.
And so, on very large scales of industrial operations, a clear 
understanding could be derived about the constrained feature of 
the space within which an organisation functions. 

The theory developed here has been subjected to empirical test with
the help of data collected from an industrial organisation that is
global in character, i.e. its presence is to be seen 
and felt all over the world. This choice is dictated 
by the requirement that one would like to understand 
the global growth behaviour of a company, whose operating space is 
by definition on the largest available scale, and, therefore, the 
overall pattern of its growth would be free of local inhomogeneities. 
To this end it has been worthwhile to study the revenue 
generating capacity
and the growth of the human resouce strength of the multi-national
company, {\it IBM}. This organisation has been in
existence in its presently known form for nearly a century. Besides
this, it has spread all over the globe. So on both of these counts,
a company like {\it IBM} is perfectly suited for the present study. 
Data about its annual revenue generation, the net annual earnings
and the cumulative human resource strength, 
dating from the year $1914$, have been published on the company 
website\footnote{\tt{http://www-03.ibm.com/ibm/history}} itself .
It has been satisfying to note that, analysed according to the 
stated objectives and specifications of this work, the {\it IBM} 
data actually give a striking match with the mathematical models 
forwarded here.
Both the capacity for revenue generation and the human resource
content of {\it IBM}, over a period of more than ninety years of
the existence of the company (this long period is actually 
quite expedient for this study, since it is concerned with
the growth of an industrial organisation from its inception to
its terminal stage), show an initial phase of exponential
growth, to be followed later by saturated growth towards a
terminal state.

A growth trend of this type can be described satisfactorily by a 
logistic differential equation, usually of second-degree nonlinearity, 
as it is done to study the growth of a population~\citep{braun,stro}. 
Regarding the study of growth from industrial data, a preceding 
work~\cite{mmr} has pedagogically underlined 
the relevance of various model differential equations of increasing 
complexity. Along these lines, a generalisation of the logistic
prescription, to any arbitrary degree of nonlinearity, is being 
posited here, to follow industrial growth through time, $t$. 
Such a generalised logistic equation will read as 
\begin{equation}
\label{logis}
{\dot{\phi}}(t) =
\lambda \phi \left(1 - \eta \phi^{\alpha} \right) \,,
\end{equation}
where $\phi$ can be any relevant variable to guage the health of a 
firm (with the ``dot" on $\phi$ being its simple time derivative), 
like its annual revenue (or cumulative revenue growth) and human 
resource strength. The parameters $\alpha$ and $\eta$ are, respectively, 
the nonlinear saturation exponent, and the ``tuning" parameter for 
nonlinearity.  Both influence the saturation behaviour of firm growth. 
In the context of the growth of an industrial organisation, one can
identify various factors that may contribute collectively to this
saturation in growth. 
These factors can be both economic and non-economic in nature. Some
may operate internally, while others can make their impact externally.
A primary factor, regarding this study at least, 
is the space within which an organisation can be
allowed to grow. If this space is constrained to be of a finite size
(as, in a practical sense, it has to be), then, of course, terminal
behaviour becomes a distinct possibility. Even as it continues to grow, 
an organisation will gradually have to contend with the boundaries
of the space within which it has to operate. This brings growth to 
a slow halt. Indeed, saturation in growth due to finite-size effects 
is understood well by now in other situations of economic 
interest where physical models can be applied~\cite{manstan}. 
The adverse conditions against growth can be further aggravated by 
the presence of rival organisations competing for the same space. 
The last factor can become particularly crucial when a miscalculation 
is made in assessing future directions of growth vis-a-vis those of 
rival organisations --- both the existing ones and the ones that
might emerge in the future.

\begin{figure}
\begin{center}
\includegraphics[scale=0.65, angle=0]{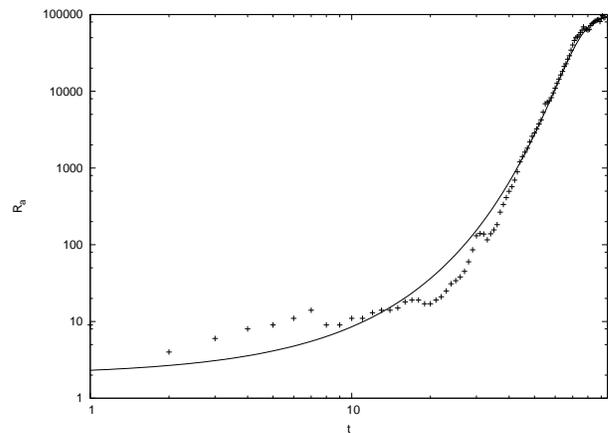}
\caption{\label{f1}\small{The continuous curve gives the model fit for
the annual revenue, $R_{\mathrm a}$, 
generated by {\it IBM}. The fit by the theoretical 
model agrees well on nonlinear time scales, for $\alpha = 1$,
$\lambda = 0.145$ and $\eta = 10^{-5}$. Here $R_{\mathrm a}$ has been 
scaled in millions of dollars, while $t$ has been scaled in years.}}
\end{center}
\end{figure}

\begin{figure}
\begin{center}
\includegraphics[scale=0.65, angle=0]{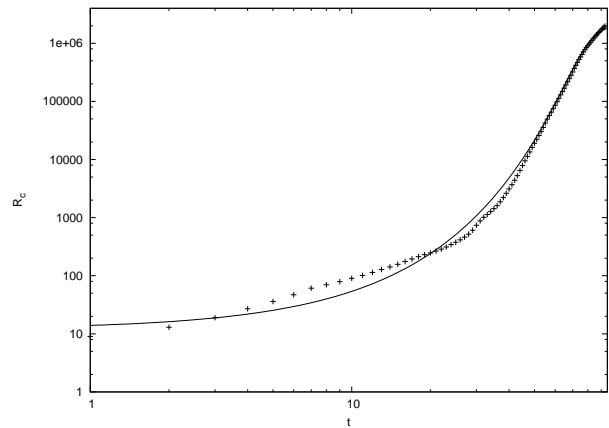}
\caption{\label{f2}\small{The {\em cumulative} growth of the annual 
revenue, $R_{\mathrm c}$, 
generated by {\it IBM} is fitted by the theoretical model (given 
as the continuous curve). The fit is very close once again on nonlinear 
scales, for $\alpha = 1$, $\lambda = 0.15$ and $\eta = 5 \times 10^{-7}$. 
As before, $R_{\mathrm c}$ has been measured in millions of dollars, 
while $t$ has been scaled in years. This distribution is evidently 
more free of fluctuations than the previous one.}}
\end{center}
\end{figure}

Integration of Eq.~(\ref{logis}), which is 
a nonlinear differential equation, yields the general integral solution 
(for $\alpha \neq 0$),
\begin{equation}
\label{integ}
\phi (t) = \left[\eta + c^{-\alpha} \exp \left(- \alpha 
\lambda t \right)\right]^{-1/\alpha} \,,
\end{equation}
in which $c$ is an integration constant. 
The fit of the foregoing integral solution with the data has been 
shown in Fig.~\ref{f1}, which gives a $\log$-$\log$ plot of the annual 
revenue, $R_{\mathrm a}$, that {\it IBM} has generated over 
time, $t$. Here the annual revenue has been measured in millions of 
dollars, and time has been scaled in years.
The data and the theoretical model given by Eq.~(\ref{integ}) agree 
well with each other, especially on mature time scales, where nonlinear 
saturation is most conspicuous. Similar features are evident in 
Fig.~\ref{f2}, which plots the {\em cumulative} growth of the 
{\it IBM} revenue, $R_{\mathrm c}$ (in millions of dollars), with 
each successive year, plotted as $t$.
The prescription of the cumulative growth needs
to be stressed upon here. For any evolving system, its rate of growth
is almost always a direct mathematical function of its current state
(which has been built up in a cumulative fashion). Very frequently 
this kind of dependence leads to an exponential growth pattern,
something that is quite relevant here, especially as regards the early 
growth of {\it IBM}. Besides, the distribution of the cumulative 
revenue is much more free of fluctuations than the distribution 
of the annual revenue, and as such the former is better suited for 
modelling. 

While the exponential feature may be appropriate for modelling
the early stages of growth, the later stages
shift into a saturation mode. A limiting value in relation to this 
saturated state could be found by setting $\dot{\phi} =0$  
in Eq.~(\ref{logis}), and this will lead to 
\begin{equation}
\label{satphi}
\phi_{\mrm{sat}} = \eta^{-1/\alpha} \,.
\end{equation}
Making use of the values of $\alpha$ and $\eta$, which have 
fitted the saturation 
properties of the plot in Fig.~\ref{f1}, a prediction can be
made that the maximum possible {\em annual} revenue ({\em not}
the cumulative revenue plotted in Fig.~\ref{f2}) that {\it IBM}
can generate will be about $100$ billion dollars.  

Another point of great interest is that the growth data 
have been fitted very well 
by the simplest possible case of nonlinearity, given by $\alpha =1$. 
This will immediately place the present mathematical problem in the 
same class of the logistic differential equation devised by Verhulst
to study population dynamics~\cite{braun,stro}. This equation has also 
been applied satisfactorily to a wide range of other cases involving 
growth~\cite{braun}. That the industrial growth model, based on the 
{\it IBM} data, should also fall within the same category is worthy
of special note. 

\begin{figure}
\begin{center}
\includegraphics[scale=0.65, angle=0]{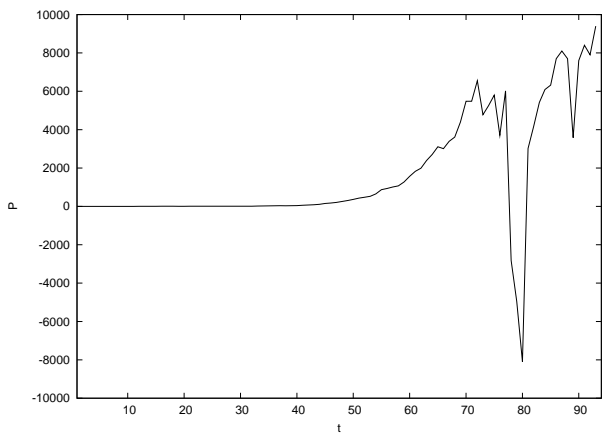}
\caption{\label{f3}\small{The net annual earnings made by {\it IBM}
has shown steady growth, except for the early years of the $1990$s
decade, which was about $80$ years of the existence of the company.
Around this time the company suffered major losses in its net earnings,
and this time scale corresponds very closely to the time scale for the
onset of nonlinear saturation in growth indicated by Eq.~(\ref{tnonlin}).}}
\end{center}
\end{figure}

The time scale for the onset of nonlinearity  
can be determined by requiring the two terms on the 
right hand side of Eq.~(\ref{integ}) to be in rough equipartition
with each other. This will give the nonlinear time scale, 
\begin{equation}
\label{tnonlin}
t_{\rm{nl}} \sim -\left({\alpha \lambda}\right)^{-1} \ln \vert \eta
c^{\alpha} \vert \,,
\end{equation} 
from which, making use of the values of $\alpha$, $\eta$ and $c$ needed
to calibrate the {\it IBM} revenue data (both the annual revenue and the
cumulative revenue), one gets $t_{\rm{nl}} \sim 75 - 80 \,{\mrm{years}}$. 
It is easy to see that this is the same time scale on which nonlinearity 
causes the onset of saturation in the growth of the company, and an 
indirect confirmation about the validity of this time scale comes from 
the plot in Fig.~\ref{f3}, which shows the growth of the net annual 
earnings of {\it IBM} (labelled as $P$, the profit), against time, $t$. 
The company suffered major reverses in its net earnings (upto $8$
billion dollars in $1993$) around $1991$-$1993$, which was indeed 
very close to $80$ years of the company, since its inception in $1914$. 
This intriguing correspondence
between the two time scales, arrived at via two distinctly different 
paths, is arguably much more than a simple coincidence. 

\begin{figure}
\begin{center}
\includegraphics[scale=0.65, angle=0]{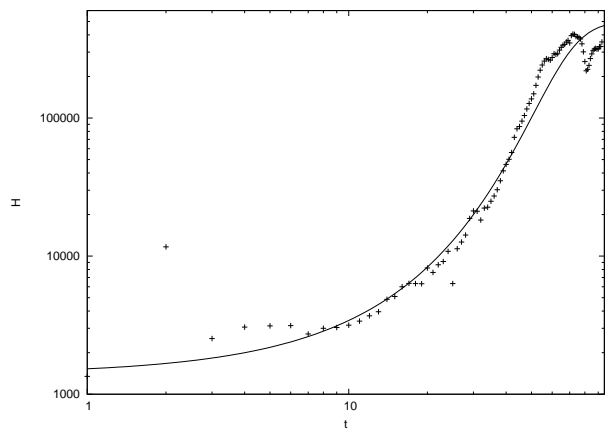}
\caption{\label{f4}\small{The growth of the human resource strength is 
fitted globally by the theoretical model for $\alpha =1$, $\lambda =0.09$ 
and $\eta = 2 \times 10^{-6}$. There has been a noticeable depletion of 
human resource on the same nonlinear saturation time scale given 
by Eq.~(\ref{tnonlin}), i.e. $75 - 80 \,{\mrm{years}}$.}}
\end{center}
\end{figure}

The human resource content of an industrial organisation is also another
indicator of its prevailing state. In the case of {\it IBM}, the data 
for the human resource of the company have been plotted in Fig.~\ref{f4}.  
Over the years the growth of human resource has been steady, and has 
followed the qualitative growth trend of the revenue. However, 
one may readily notice that just like the net earnings of {\it IBM},
there has been a sharp decline in the number of human resource around 
the time scale of $75 - 80 \,{\mrm{years}}$ of the age of the company. 
This may be construed as another independent piece of evidence in 
support of the theoretical estimate of the saturation
time scale that Eq.~(\ref{tnonlin}) indicates. And going by the values
of $\alpha$ and $\eta$ needed for the model fit in Fig.~\ref{f4}, the 
maximum possible human resource strength that {\it IBM} can viably 
employ is predicted from Eq.~(\ref{satphi}) to be about $500,000$.  

By now it has become evident that there is a strong correlation among 
the variables by which one may monitor the state of an industrial 
organisation. And in fact it happens that a complete impression 
of the true state of affairs may only be conveyed by analysing all
the variables involved. The concept of the ``Balanced Scorecard"
is somewhat related to this principle~\cite{kapnor1}. The meaningful
variables can all be diverse in nature, ranging from the finances of 
an organisation
to its human and technological resources and to the market within
which the organisation might be operating. The growth rate of any
one of these variables may have a correlated functional dependence
on the collective current state of all the pertinent variables.

So if one were to study an industrial organisation whose current state
is defined completely by a general set of $n$ variables, like 
$R_{\mathrm c}$ (or $R_{\mathrm a}$) and $H$, then the growth rate 
of the $i$-th. variable, $\phi_i$, will be given by
\begin{equation}
\label{gendyn}
{\dot{\phi_i}} = \Phi_i \left(\{\phi_j\} \right) ,
\end{equation}
with $\Phi_i$ being a general function of all the variables in the set.
The whole set can be expressed explicitly by making both $i$ and $j$
run from $1$ to $n$. This will give a set of $n$
first-order
differential equations, with each one of them being coupled to all
the others, and this entire set will form an autonomous 
first-order dynamical system in an $n$-dimensional space. 

If an industrial
organisation generates enough revenue, it becomes financially
viable for it to maintain a sizeable human resource pool. On the other
hand, the human resource strength will
translate into a greater ability to generate revenue. In this
manner both the revenue and the human resource content of an organisation
will sustain the growth of each other. Defining a general revenue 
variable,
$R$ (which can be either $R_{\mathrm a}$ or $R_{\mathrm c}$), its
coupled dynamic growth features along with the human resource, $H$, 
can be formally stated in mathematical terms as 
\begin{eqnarray}
\label{dynsys}
%\frac{{\mrm d}R}{{\mrm d}t} &=& {\rho}\left(R,H\right) \nonumber \\
\dot{R} &=& {\rho}\left(R,H\right) \nonumber \\
%\frac{{\mrm d}H}{{\mrm d}t} &=& {\sigma}\left(R,H\right) .
\dot{H} &=& {\sigma}\left(R,H\right) \,.
\end{eqnarray}
The foregoing coupled set of autonomous
first-order differential equations forms 
a two-dimensional system involving two variables, $R$ and $H$. The 
equilibrium condition of this dynamical system is obtained when both 
the derivatives on the left hand side of Eqs.~(\ref{dynsys}) vanish 
simultaneously, i.e. ${\dot R} = {\dot H} =0$. The corresponding
coordinates in the $H$---$R$ plane may be labelled $(H_0,R_0)$. Since 
the terminal state implies the cessation of all growth in time (i.e. 
all derivatives with respect to time will vanish), it is now possible 
to argue that the equilibrium state in the $H$---$R$ plane actually 
represents a terminal state in real time growth.

Some general deductions can now be made about the nature of the
equilibrium state, with the help of dynamical systems
theory~\cite{stro,js99}. The two coupled equations, given by
Eqs.~(\ref{dynsys}), will, in the most general sense,
be nonlinear. A linearisation
treatment on them could be carried out by applying small perturbations
on $R$ and $H$ about their equilibrium state values. The perturbation
scheme will be $R=R_0+R^{\prm}$ and $H=H_0+H^{\prm}$. This will allow
a coupled set of linearised equations to be set down as
\begin{eqnarray}
\label{linsys}
\frac{{\mathrm d}R^{\prime}}{{\mathrm d}t} 
&=& {\mathcal A}R^{\prime}
+ {\mathcal B}H^{\prime} \nonumber \\
\frac{{\mathrm d}H^{\prime}}{{\mathrm d}t} 
&=& {\mathcal C}R^{\prime}
+ {\mathcal D}H^{\prime} ,
\end{eqnarray}
in which
\begin{eqnarray}
\label{coeffs}
{\mathcal A} = \frac{\partial {\rho}}{\partial R}
\bigg{\vert}_{R_0} \,,\qquad & &
{\mathcal B}= \frac{\partial {\rho}}{\partial H}
\bigg{\vert}_{H_0} \,, \nonumber \\
{\mathcal C} = \frac{\partial {\sigma}}{\partial R}
\bigg{\vert}_{R_0} \,,\qquad & &
{\mathcal D}= \frac{\partial {\sigma}}{\partial H}
\bigg{\vert}_{H_0} \,.
\end{eqnarray}
Solutions of the form $R^{\prime} \sim e^{\omega t}$ and
$H^{\prime} \sim e^{\omega t}$ will enable one to derive the
eigenvalues, $\omega$, of the stability matrix implied by
Eqs.~(\ref{linsys}), as
\begin{equation}
\label{eigen}
\omega = \frac{1}{2}
\left[\left({\mathcal A}+{\mathcal D}\right)\pm
\sqrt{\left({\mathcal A}+{\mathcal D}\right)^2 -
4\left({\mathcal A}{\mathcal D} -
{\mathcal B}{\mathcal C}\right)}\right] \,.
\end{equation}
The exact determination of the values of the two roots of $\omega$
will impart a clear idea about the nature of the equilibrium state,
which can either be a saddle point or a node or a focus~\cite{stro,js99}.
The last case will necessarily mean an oscillatory nature in the
growth of both $R$ and $H$ through time~\cite{stro,js99}, and is
not tenable here. Going back to
Figs.~\ref{f1},~\ref{f2} \&~\ref{f4}, one notices that the respective 
growth patterns of {\em both} $R$ and $H$ have, on the other hand, 
been largely monotonic in nature (except for a manifestly sharp dip
in $H$ at large values of $t$). So an immediate conclusion that follows 
is that the equilibrium state is very likely a node~\cite{stro,js99}, 
and this will correspond mathematically to $\omega$ having two real 
roots of the same sign. Practically speaking, this is what is to be 
expected entirely. For an industrial organisation it is not conceivable 
that while there is growth in one variable, there will be decay in the 
other (if this were to happen, the equilibrium state will be a saddle 
point). Both 
will have to grow in close mutual association, and, if the dynamical 
systems argument is anything to go by, both will make the approach 
towards the terminal state simultaneously. The qualitative behaviour 
of one variable cannot be completely independent of the other.

\begin{figure}
\begin{center}
\includegraphics[scale=0.65, angle=0]{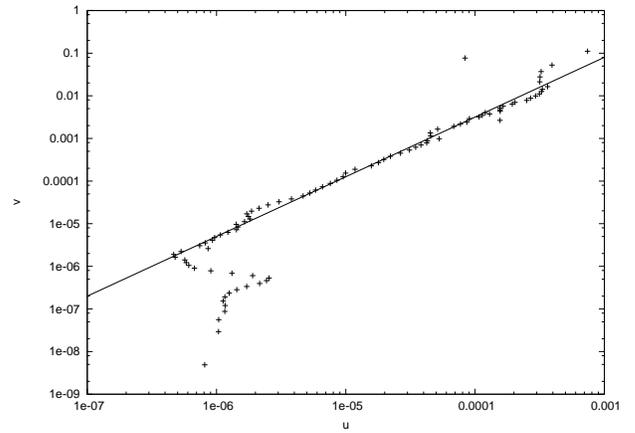}
\caption{\label{f5}\small{The straight-line fit validates the logistic
equation model. The slope of the straight line is given as $1.4$, and 
it closely matches the value of $1.6$ that could be obtained from the 
parameter fitting in Figs.~\ref{f2} \&~\ref{f4}. The cusp at the 
bottom left is due to the loss of human resource. The lower arm of 
the cusp has nearly the same slope as the straight-line above it.
Growth, corresponding to the positive slope in this plot, can be 
modelled well by the logistic equation.}}
\end{center}
\end{figure}

To have a quantitative corroboration of the dynamical systems modelling
with the help of the {\it IBM} data, one could, 
motivated by Eq.~(\ref{logis}), 
proceed with a simple ansatz to describe the individual growth patterns 
of $R$ and $H$ by an uncoupled logistic equation model. This can be 
set down as 
\begin{eqnarray}
\label{arraylogis}
{\dot{R}}(t)&=&\lambda_r R\left(1 -\eta_r R^{\alpha_r}\right)\nonumber\\
{\dot{H}}(t)&=&\lambda_h H\left(1 -\eta_h H^{\alpha_h}\right) \,,
\end{eqnarray}
with the subscripts $r$ and $h$ in the parameters $\alpha$, $\lambda$
and $\eta$, indicating that $R$ and $H$ will each, in general, have 
its own different set of parameter values. From Eqs.~(\ref{arraylogis}) 
one could easily eliminate the two time derivatives and arrive at 
\begin{equation}
\label{auton}
\frac{{\mathrm d}R}{{\mathrm d}H} = 
\frac{\lambda_r R\left(1 -\eta_r R^{\alpha_r}\right)}
{\lambda_h H\left(1 -\eta_h H^{\alpha_h}\right)} \,,
\end{equation}
whose integral solution can be expressed in a compact power-law form as
\begin{equation}
\label{powlaw}
v = \kappa u^{\beta} \,,
\end{equation} 
under the definitions that
\begin{equation}
\label{defin}
v = \frac{1}{R^{\alpha_r}} - \eta_r \,, \qquad
u = \frac{1}{H^{\alpha_h}} - \eta_h \,, \qquad
\beta = \frac{\alpha_r \lambda_r}{\alpha_h \lambda_h} \,,
\end{equation}
and $\kappa$ is an integration constant. The power law in 
Eq.~(\ref{powlaw}) implies that a $\log$-$\log$ plot 
of $v$ against $u$ will be a straight line with a slope, $\beta$. 
For the greater part of it, this fact has been established 
emphatically in Fig.~\ref{f5}. 
In this plot, $v$ has been defined in terms of the cumulative
revenue, i.e. $R=R_{\mathrm c}$, while going by the model fitting
in Figs.~\ref{f1},~\ref{f2} \&~\ref{f4}, the values 
$\alpha_r = \alpha_h =1$, $\eta_r = 5 \times 10^{-7}$ and 
$\eta_h = 2 \times 10^{-6}$ have been used. The cusp in the bottom 
left corner of the plot has arisen because of an irregular depletion 
of human resource in {\it IBM} in the early $1990$s, and hence it is 
beyond the scope of the logistic equation model. However, the lower 
arm of the cusp has nearly the same positive slope as the 
straight-line fit. So, as long as there is steady growth in both 
$R$ and $H$, the logistic equation model gives a good description 
of the global performance of an industrial organisation. A company
can be said to be performing well if it is on a curve with a positive
slope in the $u$---$v$ plot, as it has been implied in Fig.~\ref{f5}.
Sustained deviations from the straight line and having a negative
slope in this $\log$-$\log$ 
plot, are causes for worry regarding the well-being of the company.
The slope of the straight line in Fig.~\ref{f5} sets a value for the 
power-law exponent in Eq.~(\ref{powlaw}) as $\beta \simeq 1.4$, 
which is quite close to the value of $\beta \simeq 1.6$, found simply 
by taking the ratio of the respective theoretical values of $\lambda$, 
chosen to fit the empirical data in Figs.~\ref{f2} \&~\ref{f4}. 

It is remarkable that the simple logistic equation employed for the 
dynamical system in Eqs.~(\ref{arraylogis}), would suffice to model 
the {\it IBM} data so closely, as Fig.~\ref{f5} shows in no uncertain
a manner. Once it is obvious
that the dynamical systems modelling has been successful, one could
formulate a precise description of the terminal state. This can be 
done by referring to Eqs.~(\ref{coeffs}) and~(\ref{eigen}), for 
which, going by Eqs.~(\ref{arraylogis}), it is easy to see that
neither does $\rho$ depend on $H$, nor does $\sigma$ depend on $R$. 
Consequently, one has ${\mathcal B} = {\mathcal C} =0$. Immediately
the two roots of Eq.~(\ref{eigen}) are found to be 
$\omega ={\mathcal A}$ and $\omega ={\mathcal D}$. To ascertain the
exact values of $\mathcal A$ and $\mathcal D$ in terms of the known
parameters, one will have to know the equilibrium 
values of $R$ and $H$. These can be shown to be $R_0 = \eta_r^{-1}$
and $H_0 = \eta_h^{-1}$, corresponding to $u=v=0$ for the terminal 
state. After this, it is straightforward to argue
that ${\mathcal A} = - \lambda_r$ and ${\mathcal B} = -\lambda_h$.
Since the values of $\lambda$ are positive in all the cases of 
model fitting, it is now possible to claim that with both the roots
of $\omega$ being real negative numbers, the limiting state for
industrial growth is represented by a stable node in the phase 
portrait of an autonomous first-order dynamical system~\cite{js99}. 
Extending 
this contention further, the limiting state can be perceived to be 
an attractor state, towards which there will be an asymptotic approach 
through an infinite passage of time~\cite{js99}. At 
least this is what the growth data pertaining to {\it IBM} indicate. 

The terminal feature in industrial growth derives from the nonlinear 
term in the general logistic equation used for the modelling. 
This will naturally suggest that the retarding factors acting against
the growth of an organisation are nonlinear in character. These
factors are all quantified through the parameters $\alpha$ and 
$\eta$ in Eq.~(\ref{logis}). Of these two parameters,
$\eta$ is more amenable to manipulations, and  
conditions conducive to lasting growth will
necessitate tuning it down. This 
ought to be the guiding principle behind a successful management 
strategy for feasible long-term growth, especially in the case of
organisations that are still in their early stages. Knowledge 
of the general nature of the possible adversities lying ahead, 
can enable a company to apply appropriately corrective measures 
at the right juncture, and defer the arrival of the nonlinear time 
scale. As a result it will make a more effective implementation of 
future strategies and innovative solutions for growth, all of which 
should be in a state of adaptable alignment with their objectives
and core competencies. The ``Blue Ocean" strategy~\cite{kimmau}, 
for instance, might be one such practical and effective means. 

On the other hand, it would never be too late for companies which 
would already have entered a terminal phase, to institute proper
reality checks. This will allow for a more functional
and timely redefining of fundamental objectives. The solutions which
might follow, could be varied in many unexpected ways. Rather than
adhering to conventional modes of growth and survival, industries
could devise ways of preserving both their existence and their
relevance by being more integrated with the social welfare of their
markets --- in short, assist 
%(perhaps even, in a partial measure at least, assume roles which 
%hitherto were largely the preserve of the state) 
in creating an environment of overall prosperity and a general
feeling of well-being, right alongside the creation of wealth. 
The benefits derived through such inclusive strategy implementation 
should be lasting. 

\begin{acknowledgments}
The author thanks A. Basu, J. K. Bhattacharjee, S. Bhattacharya, 
B. K. Chakrabarti, I. Dutta, T. Ghose, A. Kumar, A. Marjit, 
S. Marjit, H. Singharay and J. Spohrer for useful comments.  
Much help came from A. Varkey in collecting data. 
\end{acknowledgments}

\bibliography{ibm_ms}

\begin{thebibliography}{9}
\expandafter\ifx\csname natexlab\endcsname\relax\def\natexlab#1{#1}\fi
\expandafter\ifx\csname bibnamefont\endcsname\relax
  \def\bibnamefont#1{#1}\fi
\expandafter\ifx\csname bibfnamefont\endcsname\relax
  \def\bibfnamefont#1{#1}\fi
\expandafter\ifx\csname citenamefont\endcsname\relax
  \def\citenamefont#1{#1}\fi
\expandafter\ifx\csname url\endcsname\relax
  \def\url#1{\texttt{#1}}\fi
\expandafter\ifx\csname urlprefix\endcsname\relax\def\urlprefix{URL }\fi
\providecommand{\bibinfo}[2]{#2}
\providecommand{\eprint}[2][]{\url{#2}}

\bibitem[{\citenamefont{Bouchaud}(2008)}]{bouch}
\bibinfo{author}{\bibfnamefont{J.-P.} \bibnamefont{Bouchaud}},
  \bibinfo{journal}{Nature} \textbf{\bibinfo{volume}{455}},
  \bibinfo{pages}{1181} (\bibinfo{year}{2008}).

\bibitem[{\citenamefont{Braun}(1978)}]{braun}
\bibinfo{author}{\bibfnamefont{M.}~\bibnamefont{Braun}},
  \emph{\bibinfo{title}{Differential Equations and Their Applications}}
  (\bibinfo{publisher}{Springer-Verlag}, \bibinfo{address}{New York},
  \bibinfo{year}{1978}).

\bibitem[{\citenamefont{Strogatz}(1994)}]{stro}
\bibinfo{author}{\bibfnamefont{S.~H.} \bibnamefont{Strogatz}},
  \emph{\bibinfo{title}{Nonlinear Dynamics and Chaos}}
  (\bibinfo{publisher}{Addison-Wesley Publishing Company},
  \bibinfo{address}{Reading, MA}, \bibinfo{year}{1994}).

\bibitem[{\citenamefont{Jordan and Smith}(1999)}]{js99}
\bibinfo{author}{\bibfnamefont{D.~W.} \bibnamefont{Jordan}} \bibnamefont{and}
  \bibinfo{author}{\bibfnamefont{P.}~\bibnamefont{Smith}},
  \emph{\bibinfo{title}{Nonlinear Ordinary Differential Equations}}
  (\bibinfo{publisher}{Oxford University Press}, \bibinfo{address}{Oxford},
  \bibinfo{year}{1999}).

\bibitem[{\citenamefont{Aghion and Howitt}(1998)}]{aghow}
\bibinfo{author}{\bibfnamefont{P.}~\bibnamefont{Aghion}} \bibnamefont{and}
  \bibinfo{author}{\bibfnamefont{P.}~\bibnamefont{Howitt}},
  \emph{\bibinfo{title}{Endogenous Growth Theory}} (\bibinfo{publisher}{The MIT
  Press}, \bibinfo{address}{Cambridge, Massachusetts}, \bibinfo{year}{1998}).

\bibitem[{\citenamefont{Marjit et~al.}()\citenamefont{Marjit, Marjit, and
  Ray}}]{mmr}
\bibinfo{author}{\bibfnamefont{A.}~\bibnamefont{Marjit}},
  \bibinfo{author}{\bibfnamefont{S.}~\bibnamefont{Marjit}}, \bibnamefont{and}
  \bibinfo{author}{\bibfnamefont{A.~K.} \bibnamefont{Ray}},
  \eprint{arXiv:0708.3467}.

\bibitem[{\citenamefont{Mantegna and Stanley}(2000)}]{manstan}
\bibinfo{author}{\bibfnamefont{R.~N.} \bibnamefont{Mantegna}} \bibnamefont{and}
  \bibinfo{author}{\bibfnamefont{H.~E.} \bibnamefont{Stanley}},
  \emph{\bibinfo{title}{An Introduction to Econophysics}}
  (\bibinfo{publisher}{Cambridge University Press},
  \bibinfo{address}{Cambridge}, \bibinfo{year}{2000}).

\bibitem[{\citenamefont{Kaplan and Norton}(1996)}]{kapnor1}
\bibinfo{author}{\bibfnamefont{R.~S.} \bibnamefont{Kaplan}} \bibnamefont{and}
  \bibinfo{author}{\bibfnamefont{D.~P.} \bibnamefont{Norton}},
  \emph{\bibinfo{title}{The Balanced Scorecard}} (\bibinfo{publisher}{Harvard
  Business School Press}, \bibinfo{address}{Boston}, \bibinfo{year}{1996}).

\bibitem[{\citenamefont{Kim and Mauborgne}(2005)}]{kimmau}
\bibinfo{author}{\bibfnamefont{W.~C.} \bibnamefont{Kim}} \bibnamefont{and}
  \bibinfo{author}{\bibfnamefont{R.}~\bibnamefont{Mauborgne}},
  \emph{\bibinfo{title}{Blue Ocean Strategy}} (\bibinfo{publisher}{Harvard
  Business School Press}, \bibinfo{address}{Boston}, \bibinfo{year}{2005}).

\end{thebibliography}

\end{document}